\documentclass[english,aps,prl,twocolumn,superscriptaddress,showpacs, amsmath]{revtex4-1}

\usepackage[T1]{fontenc}
\usepackage[latin9]{inputenc}
\usepackage{pdfpages}
\usepackage{color}
\usepackage{units}
\usepackage{amssymb}
\usepackage{graphicx}
\usepackage{esint}
\usepackage{bm}
\usepackage{natbib}
\usepackage{amsmath}

\usepackage{braket}
\usepackage{mathtools}
\usepackage{physics}
\usepackage{soul}

\usepackage{ulem}
\setcitestyle{journalcolor= blue}

\usepackage{babel}

\makeatletter
\AtBeginDocument{\let\LS@rot\@undefined}
\makeatother

\begin{document}
\title{Quantized conductance with non-zero shot noise as a signature of Andreev edge state}

\author{Manas Ranjan Sahu}
\thanks{equally contributed}
\affiliation{Department of Physics, Indian Institute of Science, Bangalore 560012, India}
\author{Arup Kumar Paul}
\thanks{equally contributed}
\affiliation{Department of Physics, Indian Institute of Science, Bangalore 560012, India}
\author{Jagannath Sutradhar}
\thanks{equally contributed}
\affiliation{Department of Physics, Indian Institute of Science, Bangalore 560012, India}
\author{K. Watanabe}
\affiliation{National Institute for Materials Science, 1-1 Namiki, Tsukuba 305-0044, Japan}
\author{T. Taniguchi}
\affiliation{National Institute for Materials Science, 1-1 Namiki, Tsukuba 305-0044, Japan}
\author{Vibhor Singh}
\affiliation{Department of Physics, Indian Institute of Science, Bangalore 560012, India}
\author{Subroto Mukerjee}
\affiliation{Department of Physics, Indian Institute of Science, Bangalore 560012, India}
\author{Sumilan Banerjee}
\email{sumilan@iisc.ac.in}
\affiliation{Department of Physics, Indian Institute of Science, Bangalore 560012, India}
\author{Anindya Das}
\email{anindya@iisc.ac.in}
\affiliation{Department of Physics, Indian Institute of Science, Bangalore 560012, India}

\begin{abstract}
Electrical conductance measurements have limited scope in identifying Andreev edge states (AESs), which form the basis for realizing various topological excitations in quantum Hall (QH) - superconductor (SC) junctions. To unambiguously detect AESs, we measure shot noise along with electrical conductance in a graphene-based QH-SC junction at integer filling $\nu=2$. Remarkably, we find that the Fano factor of shot noise approaches $\it{half}$ when the bias energy is less than the superconducting gap ($2\Delta$), whereas it is close to zero above $2\Delta$. This is striking, given that, at the same time, the electrical conductance remains quantized at $2e^2/h$ within and above $2\Delta$. A quantized conductance is expected to produce zero-shot noise due to its dissipationless flow. However, at a QH-SC interface, AESs carry the current in the zero-bias limit and an equal mixing of electron and hole-like states produces $\it{half}$ of the Poissonian shot noise with quantized conductance. The observed results are in accord with our detailed theoretical calculations of electrical conductance and shot noise based on non-equilibrium Green's function method in the presence of disorder. Our results pave the way in using shot noise as a detection tool in the search of exotic topological excitations in QH-SC hybrids.
\end{abstract}

\maketitle

Chiral quantum Hall (QH) edge states in proximity with a superconductor (SC) can give rise to exotic excitations \cite{beenakker2015random,bishara2007non,nayak2008non,mong2014universal,qi2010chiral,fisher1994cooper} like Majorana fermion. There are several promising theoretical proposals of realizing chiral Majorana fermion at QH-SC interfaces \cite{qi2010chiral,mong2014universal,Jose2015}, however, its evidence is still inconclusive \cite{rahmani2019interacting,manousakis2020weak,motome2020hunting,kayyalha2020absence}. Realization of the electron-hole hybrid states called Andreev edge states (AESs) at the QH-SC interface is an important step in this quest and graphene hosting clean QH edge states at moderate magnetic field is an ideal platform. The recent developments of several superconductors with large critical magnetic field 
and transparent interfaces with high quality graphene have paved the way for a number of interesting experimental observations \cite{rickhaus2012quantum,shalom2016quantum,amet2016supercurrent,park2017propagation,lee2017inducing,sahu2018inter,seredinski2019quantum,zhao2020interference,sahu2019enhanced,lee2018proximity,finkelstein2017superconductivity,dvir2021planar}, such as crossed Andreev conversion \cite{lee2017inducing}, supercurrent in QH regime \cite{amet2016supercurrent}, inter-Landau-level Andreev reflection \cite{sahu2018inter} and interference of chiral AESs \cite{zhao2020interference}. Despite these progresses, the identification of AESs remains scarce, and its dynamics have remained unexplored in the presence of disorder and dissipation.

\begin{figure}[b!]
\begin{center}
\includegraphics[width=0.5\textwidth]{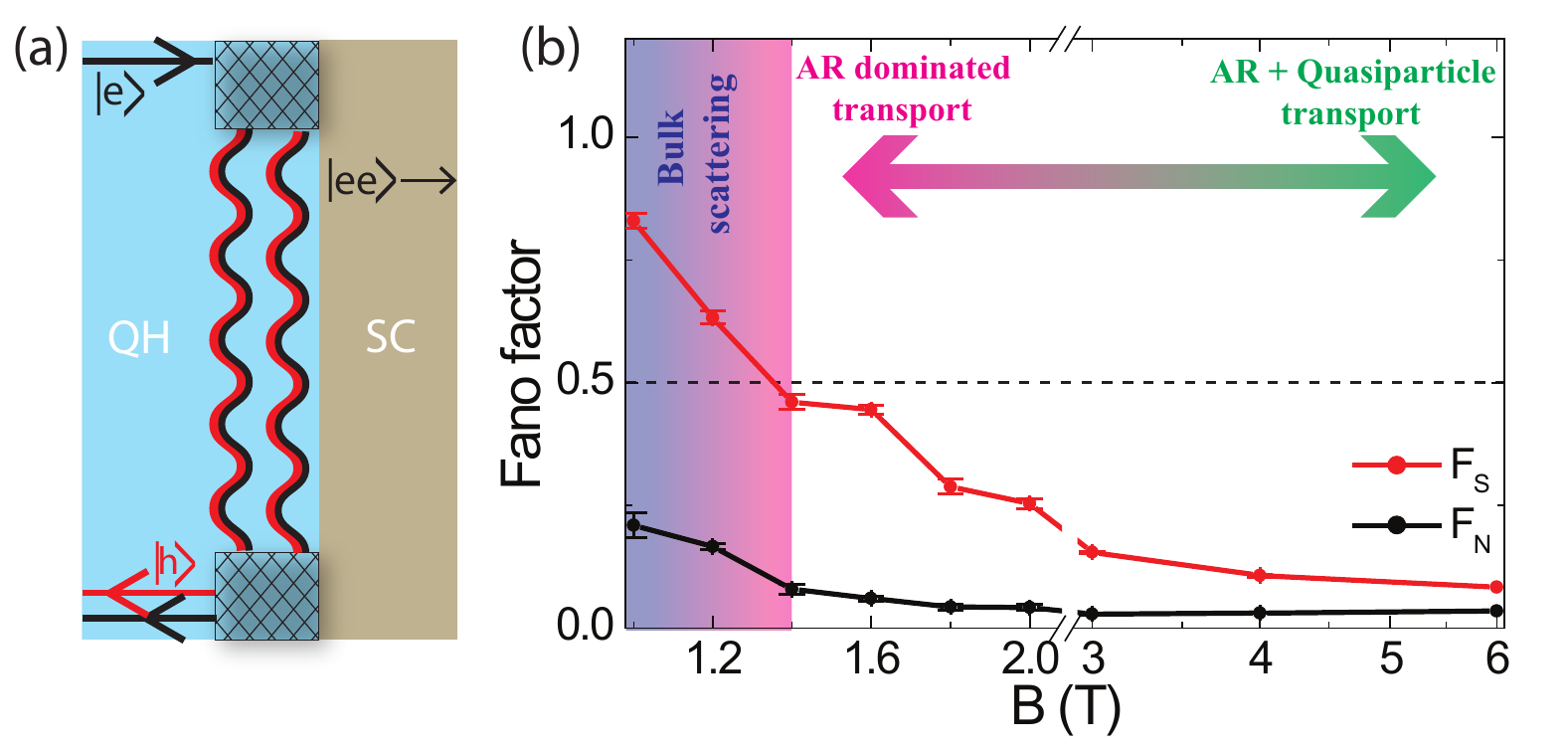}
 \caption{(\textbf{a}) Schematic of the AESs (wavy lines). (\textbf{b}) Experimentally measured low-bias Fano factor, $F_S$ and high-bias Fano factor, $F_N$ are plotted as a function of magnetic field. The different transport regimes are highlighted by the gradient colours. The dashed horizontal line correspond to $\it{half}$ Fano.
}
 \label{fig1}
 \end{center}
\end{figure} 

AESs result from the repeated Andreev reflections at a QH-SC interface, where an incident electron successively turns into a hole and back into an electron. Quantum mechanically, AESs are fermionic modes that arise from linear combinations of electron and hole, and propagate along the interface as shown schematically in Fig.~\ref{fig1}(a). The nature of the resultant fermion coming out from the interface relies on the interference of the AESs. An electron or hole or mixture of electron and hole can exit at the end of the interface depending on the phase difference accumulated by the AESs. Thus, in a conventional conductance measurement, the conductance is expected to oscillate between zero and $4e^2/h$ at  filling factor $\nu=2$ as a function of parameters like chemical potential and magnetic field, giving a robust signature of AESs. However, the inevitable presence of disorder and interface roughness randomizes the phase. As a result, low visibility of conductance oscillations around $2e^2/h$ is observed experimentally for shorter QH-SC interface\cite{zhao2020interference}. For a wider QH-SC interface, the oscillations are expected to vanish due to complete phase averaging and will exhibit quantized conductance of $2e^2/h$, exactly like a QH-normal metal (NM) junction. Hence, the conductance measurement is a limited tool to explore the AESs at QH-SC interfaces. On the contrary, as we demonstrate here, the shot noise, which originates from the discrete nature of the current carriers \cite{blanter2000shot}, does not vanish due to the phase averaging and can provide new insights into the AESs interference. The shot noise of a system is quantified by the Fano factor ($F$), namely the ratio of measured current noise with the Poissonian value of shot noise, $2eI$, for a current $I$ \cite{tworzydlo2006sub,dicarlo2008shot,danneau2008shot,kumada2015shot,matsuo2015edge,sahu2019enhanced}. Remarkably, we experimentally observe a large non-zero Fano factor, close to $\it{half}$, simultaneously with a quantized conductance in a graphene QH-SC junction, thus providing a robust and unambiguous signature of AESs. 

To this end, we perform conductance and shot noise measurement in a graphene QH-SC junction at filling factor $\nu = 2$ as a function of excitation or bias energy ($eV_{SD}$) at several magnetic fields ($B$). As shown in Fig.~\ref{fig1}(b), our shot noise results can be divided into three regions; bulk dominated transport at lower $B$, AESs dominated transport at intermediate $B$, and normal quasiparticles dominated transport at higher $B$. In the intermediate $B$, the Fano factor approaches a value close to half when the $eV_{SD} \ll 2\Delta$, whereas the Fano factor is close to zero for normal quasiparticle transport for $eV_{SD} \gg 2\Delta$ [Fig.~\ref{fig1}(b)]. In contrast the conductance always remains $2e^2/h$ across the excitation energy $2\Delta$. Our experimental observations are supported by an effective model [Fig.~\ref{fig1}(a)], discussed later, which predicts a Fano factor of half for a QH-SC junction for complete phase averaging of interference of AESs. This is in stark contrast to the zero Fano factor expected for a QH-NM junction. Moreover, we also calculate conductance and Fano factor for a microscopic model of graphene QH-SC junction in the presence of disorder using non-equilibrium Green's function (NEGF) method. We find disorder averaged conductance ($G$) with quantized plateaus ($\nu e^2/h$), as in QH-NM junction, and a large non-zero quantized Fano factor for $\nu=2$ plateau as a consequence of equal mixing of electron and hole like states.

\begin{figure}[t!]
\begin{center}
\includegraphics[width=0.5\textwidth]{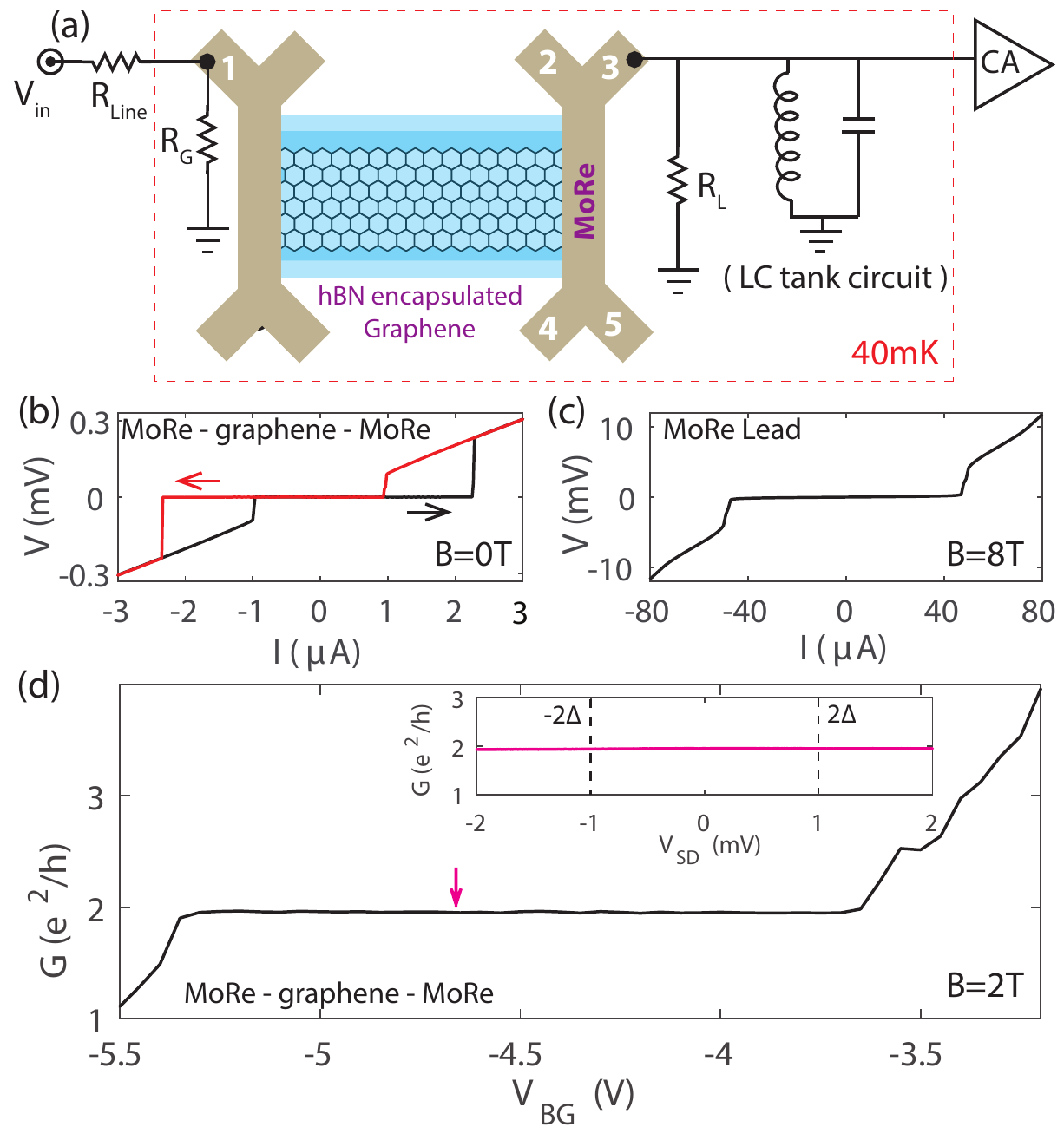}
 \caption{(\textbf{a}) The schematic of MoRe-graphene-MoRe device along with the shot noise measurement setup. (\textbf{b}) The I-V response of the MoRe-graphene-MoRe device showing supercurrent at zero magnetic field. (\textbf{c}) The I-V response of MoRe lead showing supercurrent at $B=8$T. (\textbf{d}) The conductance of the MoRe-graphene-MoRe device at $\nu =2$ QH plateau at $B=2$T plotted as a function of back gate voltage. The inset shows conductance at $\nu =2$ with bias energy. The vertical lines correspond to the superconducting gap.
}
 \label{fig2}
 \end{center}
\end{figure}


 To realize coupling of QH and superconductivity, we fabricated hexagonal Boron Nitride (hBN) encapsulated graphene device edge contacted with type-II  Molybdenum Rhenium (MoRe) superconductor on a Si/SiO$_2$ substrate. The width of the graphene-SC interface was $\sim 2~\mathrm{\mu m}$ and the channel length was $\sim 0.8~\mathrm{\mu m}$.  Fig.~\ref{fig2}(a) shows the schematic of the device with measurement setup. The device fabrication and measurement setup are discussed in details in the Supplemental Material (SM)\cite{supplement}. The MoRe leads show supercondcting transition at $T_c\sim8.7~K$ (Fig.~S5). Fig.~\ref{fig2}(b) shows the supercurrent of the MoRe-graphene-MoRe junction at zero magnetic field demonstrating the high quality of graphene-MoRe interfaces (details in SM~Sec.~S8). The supercurrent of the junction is killed by applying a tiny field of 100mT. Whereas, the MoRe leads remain superconducting at large magnetic fields as shown by the I-V characteristic in Fig.~\ref{fig2}(c) depicting supercurrent at B=8T.

Clean QH plateaus of the MoRe-graphene-MoRe junction are observed at magnetic field as low as 1T (Fig.~S6). Fig.~\ref{fig2}(d) shows two-probe conductance ($G$) around $\nu=2$ filling at $B=2$T as a function of back gate voltage ($V_{BG}$) and the conductance plateau remains very close to $2e^2/h$  similar to QH-NM interface. Further, the conductance remains almost unchanged by application of bias energy as shown in Fig.~\ref{fig2}(d)-inset. The vertical dashed lines mark the the proximity induced superconducting gap, $2\Delta\sim 1meV$ as can be seen from our shot noise data discussed later. The conductance values for $\nu=2$ and $\nu=6$ filling factors remain very close to $2e^2/h$ and $6e^2/h$, respectively, for the full range of magnetic field from 1T to 10T (Fig.~S6). These observations are in accordance with the recent experiments \cite{lee2017inducing,zhao2020interference}. However, there were no noticeable oscillations at $\nu = 2$ [Fig.~\ref{fig2}(d)] as compared to observed oscillations by Zhao et al.~\cite{zhao2020interference}.

\begin{figure}[t!]
\begin{center}
 \includegraphics[width=0.5\textwidth]{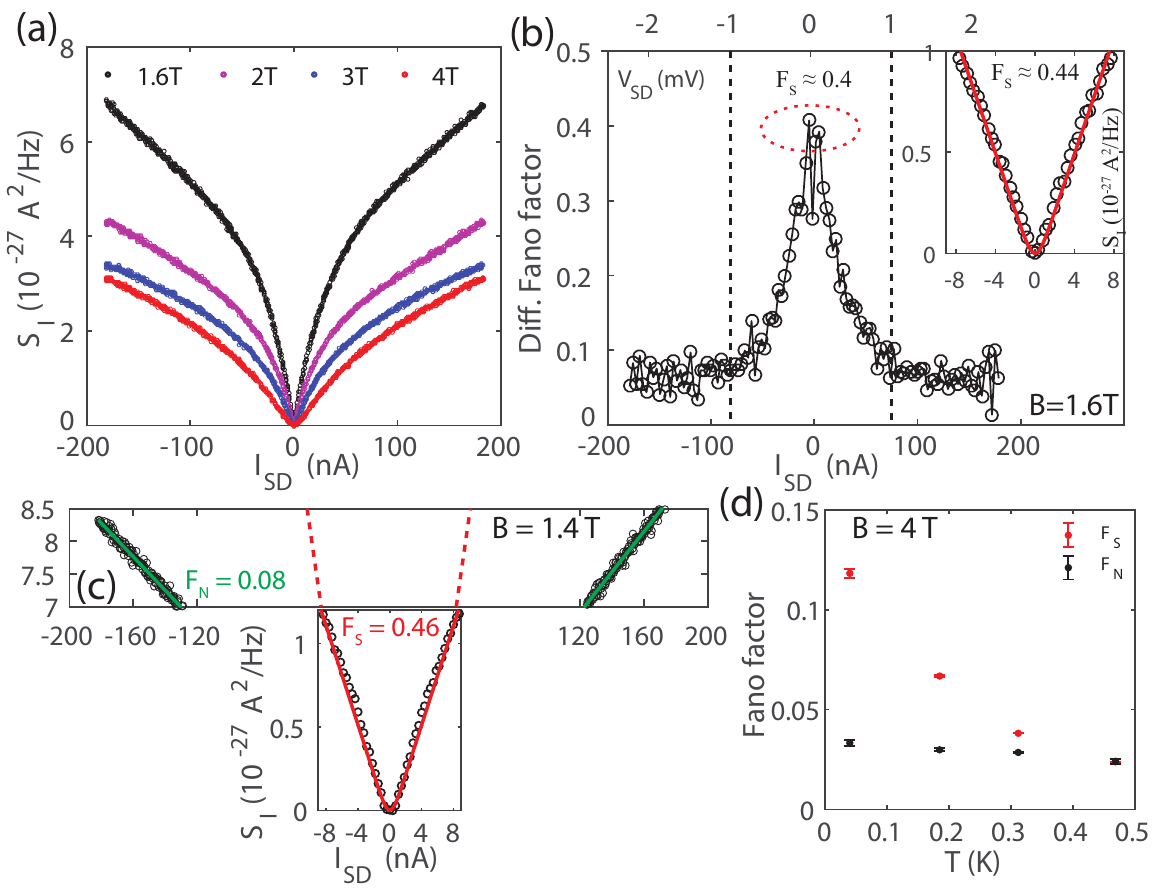}
 \caption{(\textbf{a}) The measured $S_I$ is plotted as a function of $I_{SD}$ at the center of $\nu$=2 plateau at several magnetic fields at 40 mK. (\textbf{b}) Differential Fano factor for the $\nu=2$ ($B=1.6$T) plateau plotted as a function of $I_{SD}$. Top axis shows the corresponding excitation voltage ($V_{SD}$). The two vertical dashed lines marks the approximate proximity induced superconducting gap $2\Delta \sim 1~\mathrm{meV}$. Inset shows the low bias fitting (using eq.(\ref{eq:ShotNoise})) of the same $B=1.6$T and $\nu=2$ shot noise data giving $F_S \sim 0.44$. (\textbf{c}) Low ($eV_{SD} \ll 2\Delta$) and high bias ($eV_{SD} \gg 2\Delta$) fittings are shown for the shot noise data at $\nu=2$ QH plateau at $B=1.4$T. (\textbf{d}) $F_S$ and $F_N$ for $\nu=2$ QH plateau at B=4T as a function of temperature.}
 \label{fig:fig3}
 \end{center}
\end{figure}

Now we present shot noise data ($S_I$) of the device. Fig.~\ref{fig:fig3}(a) shows the shot noise data taken at the center of the $\nu=2$ QH plateau plotted as a function of applied DC current ($I_{SD}$) at several magnetic fields. Interestingly, the $S_I$ does not increase linearly for the full range of current. Rather, a local linearity can be seen either at lower current ($< 20~\mathrm{nA}$) or at larger current ($> 100~\mathrm{nA}$) with the slope being significantly larger for the former case. For better understanding this, we plot the differential Fano factor, $F=(1/2e) (dS_I/dI_{SD})$, for $B=1.6$T as a function of $I_{SD}$ in Fig.~\ref{fig:fig3}(b). The differential Fano factor peaks around zero $I_{SD}$ and saturates to very small magnitude at larger $I_{SD}$. Such transition is expected around the superconducting gap ($2\Delta \sim 1~$meV) marked by vertical dashed lines. The observed gap is smaller than the expected BCS gap $2\Delta_{BCS} \sim 2~\mathrm{meV}$ as calculated from the critical temperature ($\sim 7-8~$K) of the bulk MoRe leads. Such reduction is observed in past and is attributed to superconducting proximity effect\cite{tikhonov2016andreev,sau2010robustness}. We define a low-bias Fano factor $F_S$ for $eV_{SD}<2\Delta$, anticipating transport via Andreev reflections. Whereas, the high-bias Fano factor for $eV_{SD}>2\Delta$ is denoted as $F_N$, which is expected to be very small since the transport happens via normal quasi-particles. From the differential Fano factor we indeed find $F_S$ to be $\sim$ 0.4 near the zero bias and $F_N$ to saturate around $\sim$0.05.

More accurate estimation of $F_S$ and $F_N$ requires fitting of $S_I$ data with the finite-temperature expression of shot noise \cite{blanter2000shot,sahu2019enhanced,paul2020interplay},
\begin{equation} \label{eq:ShotNoise}
S_I=\left\{\begin{array}{ll}
2eI_{SD}F_S \, [coth(\frac{e^*V_{SD}}{2k_BT_e})-\frac{2k_BT_e}{e^*V_{SD}}],& \text{for } |eV_{SD}|<2\Delta \\
K + 2eI_{SD}F_N,              & \text{for } |eV_{SD}|>2\Delta
\end{array}\right.\
\tag{1}
\end{equation}
Where, $V_{SD} = \frac{I_{SD}}{G}$ is the excitation voltage, $k_B$ is the Boltzmann constant and $T_e$ ($\sim 40mK$) is the electron temperature. Fig.~\ref{fig:fig3}(b)-inset shows the low bias fitting of $B=1.6$T shot noise data, which gives $F_S$ to be $\sim 0.44$. We extracted Fano factors for different magnetic fields by fitting the corresponding shot noise data with Eq.\ref{eq:ShotNoise} as shown in Fig.~\ref{fig:fig3}(c) for $B = 1.4T$. The extracted $F_S$ and $F_N$ are plotted as a function of $B$ in Fig.~\ref{fig1}(b), where it can be seen that $F_N$ remains almost constant ($\sim 0.05$) at magnetic field as low as 1.4T to high magnetic field. However, it increases  rapidly below 1.4T due to the bulk contribution. The presence of bulk contribution below $1.4$T is further justified from the rapid degradation of quality of $\nu=2$ QH plateau in this regime, which is shown in Fig.~S14. More interestingly, $F_S$ has larger value and approaches close to $\it{half}$ in the range 1.4-1.6T and then decreases with increasing magnetic field. In  SM~Sec.~S5, we discuss how the presence of vortices at the QH-SC interface can enable the normal quasi-particle transport and hence reduce the value of Fano factor with increasing magnetic field. In Fig.~\ref{fig:fig3}(d), we show the $F_S$ and $F_N$ for $\nu=2$ QH plateau with temperature at B=4T. It can be seen that $F_S$ decreases with increasing temperature and becomes almost equal to $F_N$ at $\sim0.5K$ due to the normal quasiparticle transport. The $T\sim0.5K$ is much smaller than the $T_C \sim 7-8K$ of bulk MoRe lead. This reduction of temperature is possibly due to superconducting proximity effect and is discussed in details in SM~Sec.~S9.
 Note that, for normal transport via QH edge states no shot noise is expected as the transmission probability equals to one. For a junction with reduced transparency ($T_N<1$) shot noise can be non-zero with $F_N=1-T_N$. From our conductance value for $\nu=2$ QH plateau, the $T_N \sim 0.98$. The observed $F_N \sim 0.05$ is close to the theoretically expected Fano factor $\sim 0.02$. The slight deviation at smaller magnetic fields ($1.4-1.8T$) could be due to small bulk contribution at large applied bias ($\sim2$meV), which does not affect $F_S$ measured at very small bias energy ($< 0.2$meV).

Next, we discuss an effective model for AESs as well as microscopic NEGF calculations to understand the observation of quantized conductance with large non-zero Fano factor in graphene QH-SC junction.

\textbf{An effective model for AESs:} An effective model of QH-SC interface is schematically shown in Fig.~\ref{fig1}(a)\cite{khaymovich2010andreev,zhao2020interference}. The zero-energy eigenstates of the AESs can be written in the electron-hole basis $\{\ket{e},\ket{h}\}$ as $\ket{\psi_1}=\alpha\ket{e}+\beta\ket{h}$ and $\ket{\psi_2}=\beta^*\ket{e}-\alpha^*\ket{h}$ with wave vectors $k_1$ and $k_2$ along the edge and $|\alpha|^2+|\beta|^2 = 1$. By writing an incoming electron-like state as a linear combination of AES eigenstates, the Andreev reflection probability ($P_\mathrm{AR}$) is obtained as $P_\mathrm{AR} = 4|\alpha|^2|\beta|^2\sin^2(\phi/2)$, where $\phi=(k_1-k_2)L$ is the phase difference acquired between the two AES eigenstates over the length $L$ (SM~Sec.~S1). When the AESs are neutral electron-hole hybrids, i.e. $|\alpha|^2=|\beta|^2=1/2$, $P_\mathrm{AR}(\phi)= \sin^2(\phi/2)$. For $\phi=0,2\pi$, no Cooper pair is transmitted due to the complete reflection of all the incident electrons. In the opposite limit, $\phi=\pi$, all the incident electrons transmit as Cooper pair due to perfect Andreev reflection. For $\nu=2$ filling factor, the conductance is zero in the former limit, and twice the normal state conductance, $4e^2/h$, in the later. In general, the junction conductance oscillates between these two limits if the phase difference is tuned either by carrier density or by magnetic field. For a realistic junction, the presence of disorder and inelastic processes at the current carrying edge \cite{marguerite2019imaging}, will introduce dephasing and decoherence, making the phase differences random. For a complete phase averaging of the transmitted current with uniform distribution of phase, the conductance of QH-SC junction becomes exactly same as that of QH-NM junction i.e $2e^2/h$. Moreover, for a given phase difference, $\phi$, the power spectral density of shot noise ($S_I$) \cite{blanter2000shot} can be written as $S_I(\phi)=2e^*I_t[1-P_\mathrm{AR}(\phi)]=8eI\sin^2(\phi/2)[1-\sin^2(\phi/2)]$, with the Cooper pair charge, $e^*=2e$. For a complete phase averaging, $S_I=eI$. Thus, the effective model gives a Fano factor of $\it{half}$ for QH-SC junction with perfectly quantized conductance plateaus, in agreement of our experimental observations.
   
\begin{figure}[t!]
\begin{center}
\includegraphics[width=0.4\textwidth]{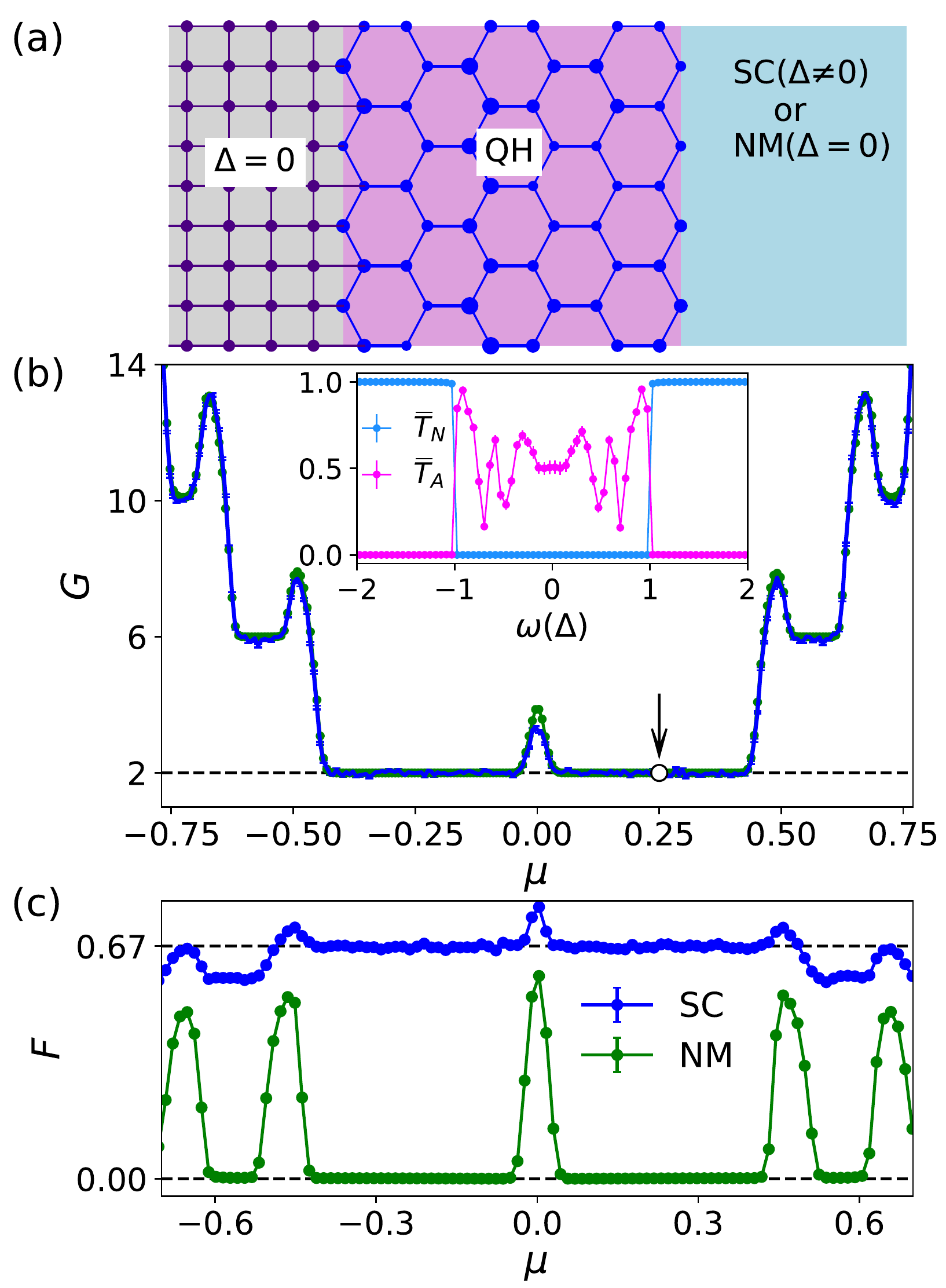}
 \caption{(\textbf{a}) System for NEGF calculations. The graphene region has a nearest neighbour hopping $t$ and magnetic field perpendicular to its plane and the varying marker size of the lattice represents a particular realization of onsite disorder. (\textbf{b}) Conductance of the system is plotted as a function of chemical potential for a disorder strength $W=0.5t$, flux $\Phi=0.07$ and SC gap $\Delta=1/2750t$ in the right lead. Conductance plateaus (blue) are realized by averaging over disorder and the plateaus appear at exactly the same conductance (green) expected for QH-NM case. \textbf{Inset:} the normal and Andreev transmission coefficient for QH-SC are plotted as a function of excitation energy at the chemical potentials highlighted by the arrow. (\textbf{c}) Calculated Fano factor is plotted as a function of $\mu$ showing the 2/3 Fano factor at the $\nu=2$ plateau. The Fano factor vanishes for QH-NM case.}
 \label{Th-fig1}
 \end{center}
\end{figure}

\textbf{NEGF calculations:} For the NEGF calculations, as in an earlier work \cite{sun2009quantum}, we consider an infinite strip of finite width  having a hexagonal graphene lattice connected to two leads one either side, as shown schematically in Fig.\ref{Th-fig1}(a). Details of the model, parameters and numerical calculations are given in  SM~Sec.~S2.
Within the NEGF formalism, the zero-temperature two-probe conductance (in units of $e^2/h$) in the linear response regime ($I=GV$) for small voltage bias $V$ is given by $G=2 \langle \left[\Tr(T_N)+ 2\Tr(T_A)\right]\rangle $, where $T_N$ and $T_A$ are normal and Adreev transmission matrices, respectively, for each spin species. The transmission matrices can be expressed in terms of Green's function and self energy of the leads (SM~Sec.~S2). Here $\langle..\rangle$ represents disorder averaging over many disorder realizations of $\{\epsilon_i\}$. We define the disorder averaged transmissions matrices as $\overline{T}_{N(A)} = \langle\Tr(T_{N(A)})\rangle$.

Fig.~\ref{Th-fig1}(b) shows the disorder averaged conductance as a function of chemical potential ($\mu$) with SC ($\Delta=1/2750t$ with nearest neighbour hopping $t$ in graphene) and normal ($\Delta=0$) right lead. The conductance for both the cases show the same QH plateaus, as expected from the effective edge model. But, from Fig.~\ref{Th-fig1}(b)-inset we can see that only Andreev transmission takes part in the transport, i.e, $\overline{T}_N=0$ and $\overline{T}_A\ne 0$, within the superconducting gap for the SC lead. 
Following Blanter et al.\cite{blanter2000shot}, the shot noise for normal-superconductor junction is given by  $S=(16e^3V/h)\Tr[T_A (1-T_A)]$, such that the Fano factor is $F=2\langle\Tr[T_A(1-T_A)]\rangle/\langle \Tr[T_A]\rangle$. The numerically computed Fano factor for the graphene QH-SC junction is shown in Fig.~\ref{Th-fig1}(c) and compared with that of QH-NM system. As evident, Fano factor for the conductance plateaus for QH-SC is non-zero, in contrast to the QH-NM system. In particular, $F=2/3$ for $\nu=2$ plateau, unlike $F=1/2$ obtained from the edge model. This difference in values of $F$ from the NEGF calculations and the effective model can be traced back to the difference of the distribution $\Tr T_A$ over disorder realizations for the two models.  As discussed in detail in SM~Sec.~S3, in case of NEGF calculations, we find that $\Tr T_A$ is uniformly distributed for $\nu=2$, thus producing $F=2/3$. On the contrary, in the effective model of AES in the preceding section, we assumed  uniform distribution of $\phi$ as opposed to the uniform distribution of $\sin^2(\phi/2)\propto T_A$. This difference between NEGF calculation and effective model does not affect the conductance. 

 To conclude, our result of close to half Fano factor along with quantized conductance for Andreev reflection dominated transport is an unambiguous signature of the existence of AESs at the QH-SC interface. The closer agreement of experimental results with the scenario of uniform phase averaging gives an indication of the crucial involvements both static disorder and inelastic processes along the current-carrying edge even at very low temperature \cite{marguerite2019imaging}. Getting insights into such intriguing dynamical processes for AESs would be of great importance for the eventual realization of novel excitations, e.g. Majorana and para-fermions, in various QH insulator-superconductor hybrid systems \cite{qi2010chiral,mong2014universal,Clarke2014,Alicea2016,Jose2015}.

A.D. acknowledges supports from the MHRD, Govt. of India under STARS research funding (STARS/APR2019/PS/156/FS), and also thanks Swarnajayanti Fellowship of the DST/SJF/PSA-03/2018-19.

\bibliography{references}{}

\onecolumngrid
\newpage
\thispagestyle{empty}
\mbox{}


\section*{Supplementary material}

**Section S2 | Model for graphene QH-SC junction and non-equilibrium Green's function (NEGF) calculation**

The system we considered in Fig. 4(a) (also shown in Fig. S1) consists of three regions (1) a hexagonal graphene lattice of finite length ($L_x$) coupled on the right with, (2) a continuous lead, either a normal metal or a superconductor, and on the left with (3) a semi-infinite metallic square lattice. The entire system is modelled by the Hamiltonian

$$H = H_L + H_G + H_R + H_{LG} + H_{RG}, \qquad (3)$$

where the Hamiltonian for the graphene part, $H_G$, is the standard nearest-neighbor tight binding model with an uniform magnetic field and given by

$$H_G = \sum_{\langle ij \rangle, \sigma} \left( t e^{i\phi_{ij}} c_{i\sigma}^\dagger c_{j\sigma} + \text{h.c.} \right) + \sum_{i,\sigma} (\epsilon_i - \mu) c_{i\sigma}^\dagger c_{i\sigma}, \qquad (4)$$

$c_{i\sigma}$ ($c_{i\sigma}^\dagger$) is the electron annihilation (creation) operator for spin $\sigma$ at the $i$-th site of the graphene hexagonal lattice; $t(=1)$ is the nearest neighbour hopping, $\phi_{ij} = \int_i^j \boldsymbol{A} \cdot d\boldsymbol{r}$ and $\boldsymbol{A}$ is the vector potential $(0, -Bx, 0)$ generating magnetic field $\boldsymbol{B} = B\hat{z}$ perpendicular to the system. $\epsilon_i$ is the onsite disorder, $\epsilon_i \in [-W/2, W/2]$. The leads and the graphene region are kept at a chemical potential $\mu$, corresponding to the middle of the SC gap, relative to the Dirac point. Thus $\mu$ controls the electron density in graphene. We have considered the lattice spacing of graphene $a = 1$ and flux $\Phi$ per plaquette is defined as $B = \dfrac{4\Phi}{3\sqrt{3}}$. $L_x$ is defined as the length of graphene along $x$-axis such that number of atoms along the top or bottom edge is $2L_x$ and $L_y$ is defined as the number of hexagon along the interface with the right lead, see Fig. S1. The results are shown for $L_x = 16$ and $L_y = 60$. The Hamiltonian for the left lead which is a tight-binding model on a semi-infinite square lattice is given by

$$H_L = \sum_{\langle \alpha\beta \rangle, \sigma} t a_{\alpha\sigma}^\dagger a_{\beta\sigma} + \text{h.c.} - \mu \sum_{\alpha, \sigma} a_{\alpha\sigma}^\dagger a_{\alpha\sigma}, \qquad (5)$$

where $a_\alpha$ is electron operator on the $\alpha$-th site of the lead. For the superconducting lead on the right, we take a continuous system instead of a lattice, i.e.

$$H_R = \sum_{\mathbf{k}, \sigma} \epsilon(\mathbf{k}) b_{\mathbf{k}\sigma}^\dagger b_{\mathbf{k}, \sigma} + \sum_{\mathbf{k}} \Delta (b_{\mathbf{k}\uparrow}^\dagger b_{-\mathbf{k}\downarrow}^\dagger + b_{-\mathbf{k}\downarrow} b_{\mathbf{k}\uparrow}), \qquad (6)$$

$b_{\mathbf{k}\sigma}$ ($b_{\mathbf{k}\sigma}^\dagger$) is the annihilates (creates) a particle of momentum $\mathbf{k} = (k_x, k_y)$ with spin $\sigma$. $\epsilon(\mathbf{k})$ follows a parabolic dispersion $\epsilon(\mathbf{k}) \propto k^2$, and $2\Delta$ is the superconducting gap. We have verified that our results remain qualitatively same for a square lattice SC lead. The couplings between the leads and graphene are given by

$$H_{LG} = \sum_{\langle ij \rangle, \sigma} t c_{i\sigma}^\dagger a_{\alpha\sigma} + \text{h.c.}, \qquad (7)$$

$$H_{RG} = \sum_{i,\sigma} t c_{i\sigma}^\dagger b_\sigma(y_j) + \text{h.c.}. \qquad (8)$$

$b_\sigma(y_i) = \sum_{k_x,k_y} e^{ik_y y_i} b_{\boldsymbol{k}\sigma}$ annihilates a particle at $y_i$ which is the vertical coordinate of the $i$-th lattice point along the interface.

**Conductance of the system:**

We compute the current at the left lead using standard NEGF formalism [5]. The total current is obtained from the currents of $\sigma = \uparrow, \downarrow$ components, $I = I_\uparrow + I_\downarrow$ where

$$I_\uparrow = \frac{e}{h} \int d\omega \, \text{Tr}\{\Gamma_{L\uparrow}[G^r \Gamma_R G^a]_{11}(f_\uparrow - f) + \Gamma_{L\uparrow} G^r_{12} \Gamma_{L\downarrow} G^a_{21}(f_\uparrow - f_\downarrow)\}, \tag{9a}$$

$$I_\downarrow = -\frac{e}{h} \int d\omega \, \text{Tr}\{\Gamma_{L\downarrow}[G^r \Gamma_R G^a]_{22}(f_\downarrow - f) + \Gamma_{L\downarrow} G^r_{21} \Gamma_{L\uparrow} G^a_{12}(f_\downarrow - f_\uparrow)\}. \tag{9b}$$

In our case, $I_\uparrow = I_\downarrow$. In the above, $G^{r(a)}$ denotes retarded (advanced) Green's function in the $2 \times 2$ particle-hole Nambu space, and $\Gamma_{l\sigma} = i(\Sigma^r_{l\sigma} - \Sigma^a_{l\sigma})$ ($l = L, R$), are obtained from the self-energies of the graphene due to the leads. $f(\omega)$ is the Fermi function. For a voltage difference $V$ between the leads $f_\uparrow(\omega) = f(\omega - eV)$ and $f_\downarrow(\omega) = f(\omega + eV)$. We obtain, $I_\uparrow = \frac{e^2}{h}(T_{N\uparrow} + 2T_{A\uparrow})V$, where the normal transmission coefficient is given by

$$T_{N\uparrow} = \Gamma_{L\uparrow}[G^r \Gamma_R G^a]_{11} \tag{10}$$

and Andreev transmission by

$$T_{A\uparrow} = \Gamma_{L\uparrow} G^r_{12} \Gamma_{L\downarrow} G^a_{21}. \tag{11}$$

Similarly, $T_{N\downarrow}$ and $T_{A\downarrow}$ can be defined for the $I_\downarrow$ and they will be equal to $T_{N\uparrow}$ and $T_{A\uparrow}$, respectively. So we define $T_N = T_{N\uparrow} = T_{N\downarrow}$ and $T_A = T_{A\uparrow} = T_{A\downarrow}$. Therefore, the conductance in unit of $e^2/h$ is

$$G = \frac{I}{V} = 2(\text{Tr}T_N + 2\text{Tr}T_A) \tag{12}$$

The factor of two with $\text{Tr}T_A$ can be attributed to the effective $2e$ charge transfer when an electron (hole) reflects back as hole (electron). Using the fact that the full $\Gamma_L$ and $\Gamma_R$ matrix have non-zero elements only at the edge of the system, we can see the components required for $T_N$ are $G^r_{1N}$ and $G^a_{N1}$, and $G^a_{N1} = (G^r_{1N})^\dagger$, where subscript 1 and $N$ represent the atomic layer of left and right interfaces of graphene, respectively. For $T_A$ we need $G^r_{11}$ and $G^a_{11}$, where $G^a_{11} = (G^r_{11})^\dagger$. We can calculate these components of Green's function in an efficient way using the standard recursive method described in Refs.6, 7. The self energy of the disorder-free semi-infinite square lattice lead can also be calculated using the same method [8]. The self energy calculation of the right continuous lead can be found in Ref.5.

**NEGF results for noise for the $\nu = 6$ QH plateau:**

Here we briefly discuss the distribution of Andreev transmission coefficient $T_A$ over disorder realizations for $\nu = 6$ conductance plateau. The results for the distribution of $\text{Tr} T_A$ and $\text{Tr}[T_A(1 - T_A)]$ related to conductance and noise, respectively, are shown in Fig. S2. As stated in the main text, and will be discussed in Sec. S3, $T_A$ is uniformly distributed for $\nu = 2$ plateau. This behaviour can be understood by looking at the eigenvalues $\{\lambda_i\}$ of $T_A$. At the first plateau, there is only one nonzero eigenvalue $\lambda_1 \in [0, 1]$ due to one edge channel for each spin component. For the $\nu = 6$ plateau $T_A$ has three nonzero eigenvalues. So, we can expect that the distribution of $\text{Tr} T_A = \sum_{i=1}^{3} \lambda_i$ can be approximated by a Gaussian, assuming central limit theorem for sum of three random numbers. We indeed find the distribution to be close to a Gaussian, but with a variance lesser than expected (Fig. S2). This is presumably due to the fact that the three eigenvalues for a single disorder realization are not completely uncorrelated. We find $\langle \text{Tr} T_A \rangle = 1.04$ and $\langle \text{Tr} T_A(1 - T_A) \rangle = 0.43$ resulting in a Fano factor $F = 0.58$.

**Section S3 | Origin of different Fano factors in the effective model and the NEGF calculation**

In case of NEGF calculations, $\text{Tr} T_A$ is uniformly distributed for $\nu = 2$ (Fig. S3), which effectively has a single conducting edge channel, and thus $T_A$ is scalar for each spin component. As a result, uniform distribution of $T_A$ implies a distribution $1/\sqrt{0.25 - x}$ for $x = \text{Tr}[T_A(1 - T_A)]$, and thus producing $F = 2/3$. From the perspective of the effective model discussed in Sec. S1, this can be understood as the uniform distribution of $\sin^2(\phi/2) \propto T_A$, as opposed to the uniform distribution of $\phi$, that was assumed earlier. This difference between NEGF calculation and the effective model does not affect the conductance. The reason behind uniform distribution of $T_A$ for $\nu = 2$ can be understood by looking at the conductance of the QH-SC junction plotted as a function of $\mu$ for a single disorder realization in Fig. S3 (c). The non-disorder-averaged conductance irregularly oscillates between 0 and 4 $e^2/h$ for QH-SC and does not exhibit any plateau, unlike that for QH-NM junction. This lack of *self-averaging property* for AES transport can be rationalized from the fact that for a fixed static disorder realization, a chiral electron entering the QH-SC interface from the upper edge [manuscript Fig. 1(a)] exits at the lower edge as a particular electron-hole superposition in a deterministic fashion, like in any static interference setup. Hence the conductance oscillates as a function of $\mu$, the details depending on the width of the junction and the particular disorder profile. This is in contrast to the effective model with complete phase averaging or a semiclassical skipping-orbit picture [9] (Fig. S4) where an incoming electron will always exit as a equal mixture of electron and hole for a wide enough junction, thus giving $G = \nu e^2/h$ plateaus. Note that, in the absence of Josephson effect, a SC-QH-SC junction will produce same results as that of a QH-SC junction[10,11] that has been considered in the effective model or for the NEGF calculations for computational convenience.

**Section S4 | Treating QH-SC junction as a set of normal ballistic modes connected to a superconductor**

Here, we consider a case, in which we treat the QH-SC junction as a ballistic system containing a finite number of modes connected to a superconductor[12]. In such a system, when the superconductor is in normal state, the expected Fano factor becomes $1 - T_N$, where $T_N$ is the transmittance of the ballistic modes. One thing is certain for our device, both from the conductance and shot noise measurements, that $T_N$ is very close to unity. In this highly transparent regime Andreev reflection probability becomes $\sim T_N^2$. This will produce a fano factor $2(1 - T_N^2)$ for the QH-SC junction. By assuming QH as a ballistic conductor, one will expect conductance doubling in the QH plateaus of the QH-SC junction compared to the QH-NM case for a highly transparent junction. The unchanged conductance of the QH plateaus of our graphene QH-SC device thus indicates that the QH-SC junction is quite different than the junction of a ballistic conductor and a superconductor. Note that, if we still use this model to calculate the Fano factor, the $T_N \sim 0.98$ extracted from the conductance measurement will produce Fano factor $\sim 0.08$ significantly smaller than the experimentally measured Fano factor, $F_S \sim 0.46$, in our device. Thus, treating the QH as a set of ballistic mode does not explain our results.

**Section S5 | Reduction of Fano factor ($F_S$) with increasing magnetic field**

We observe that $F_S$ reduces with increasing magnetic field and we attribute this to possible normal quasi-particle leaking through vortices in the type-II superconductor MoRe. How such process may not affect the conductance but can affect the shot noise can be understood qualitatively by employing a simple model[13]. Let's assume that, the probability of an incident electron at QH-SC interface will transmit as a normal quasiparticle be $\alpha$. The remaining $1 - \alpha$ fraction of electron will then pass through the AESs transmitting $(1 - \alpha)/2$ Cooper pairs into the superconductor. As we have discussed in Sec. S1 and S2, the conductance of QH-SC junction is same as the conductance of QH-NM junction, and as a consequence the conductance of junction will not be affected by non-zero normal quasiparticle leaking. But, as the significant Fano factor contribution comes from AR processes, the normal quasiparticle leak will reduce the Fano factor. With increasing magnetic field the vortex density in superconductor increases and this subsequently will increase the probability of normal quasiparticle leaking $\alpha$, thus reducing the shot noise, in qualitative agreement with our observation.

**Section S6 | A semi-classical model**

Here we present a semi-classical model of AR at QH-SC junction, which nicely captures its unchanged conductance compared to the QH-NM junction case [9]. For an ideal QH-SC junction an incident electron at one end of the interface will Andreev reflect a hole, and then after the second bounce the incident hole will

Andreev reflect an electron, and the process will go on. Due to this repetitive AR process, if an electron comes out at the other end of the junction, it means no cooper pair has been transmitted; whereas, if a hole comes out, then effectively one cooper pair has been transmitted. the first case will give zero conductance and the second case will give twice quantum conductance, and this will solely be decided by width of the junction. But in reality there is always some barrier at the interface, which reduces the probability of AR ($P_{AR}$) from unity. As a consequence, after the first bounce the probability of getting a hole will be, $P_{hole}(1) = P_{AR}$ and the probability of getting an electron will be, $P_{el}(1) = 1-P_{AR}$. As the number of bounce increases, the probability of getting a hole will be decided by a recurrence relationship,

$$P_{hole}(n+1) = P_{hole}(n) \times (1 - P_{AR}) + (1 - P_{hole}(n)) \times P_{AR}. \tag{13}$$

The probability function $P_{hole}$ takes the value of exactly half irrespective of $P_{AR}$ after several bounces as shown in Fig. S4. This co-propagating electron-hole mixed edge states at the QH-SC interface forms the Andreev edge sate. because of this equal mixing of electron and hole the conductance becomes exactly what is expected with normal contacts.

**Section S7 | Experimental methods**

**Device fabrication:** To make the device, the graphene and Hexagonal Boron nitride (hBN) flakes were exfoliated from bulk crystal on Si/Sio2 substrates. For graphene we have used HOPG crystals. The suitable monolayer graphene flake and hBN flakes were first identified under optical microscope and then stacked sequentially using PDMS - PC dry transfer technique[14–16]. The MoRe edge contacts on the hBN encapsulated graphene were realized by first defining the contact areas using standard e-beam lithography, followed by reactive ion etching ($CHF_3+O_2$) and deposition of MoRe using DC sputtering with Ar-plasma.

**Measurement technique:** Only the MoRe characterization experiments were carried out in He3 cryostat. All other experiments were performed in a dilution refrgerator having 40 mK base temperature. We employ a LC tank circuit based high frequency measurement circuit explained in detail in our previous work[12]. We have further discussed the working of shot noise measurement circuit in Fig. S7.

**Extraction of shot noise:** The measured noise ($S_V$) contains different other types of noises together with the required shot noise component ($S_I$). The total measured noise can be written as[12, 17, 18],

$$S_V = S_I R_{eff}^2 + 4k_B T R_{eff} + S_i^{amp} R_{eff}^2 + S_v^{amp}$$

Here, $R_{eff}$ is the parallel effective resistance of the sample ($R_S = 1/G$) with the load resistor ($R_L$); $R_G \sim 5\Omega$ is negligible compared to $R_S$. $4k_B T R_{eff}$ is the thermal noise, $S_i^{amp}$ and $S_v^{amp}$ are the current and

voltage noise of the cold amplifier (CA), respectively. Thus, the shot noise, which is the current dependent excess noise, can be extracted from the total noise as, $S_I(I_{SD}) = \frac{S_V(I_{SD}) - S_V(I_{SD}=0)}{R_{eff}^2}$.

**Section S8 | Device characterization at zero magnetic field**

The device shows supercurrent at all carrier densities as shown in Fig. S9(a), with highest critical current of $\sim 2.3\mu A$ (Fig. S9(b)) at high doped electron side. The hysteresis in supercurrent curves (Fig. S9(b)) is a characteristic of underdamped Josephson junctions and is frequently observed in past in graphene based Josephson junctions[19–21]. We note that electron heating effect can also give rise to such asymmetry[22–24]. The electron-hole asymmetry in supercurrent magnitude (Fig. S9(a)) is also observed in past and is attributed to contact induced n-type doping and consequent formation of pn junctions close to the contacts[20–22]. In Fig. S10(b) we have plotted the $I_C R_N$ product, which holds key information about device quality. At large carrier densities the $I_C R_N$ product remain almost independent of gate voltage indicating the device is ballistic in nature. Note that, for a diffusive system $I_C R_N$ product depends on mean free path, which in turn depends on gate voltage. Further, the $I_C R_N$ product is much smaller than the theoretically predicted value for ballistic Josephson junction $\approx 2.1\Delta/e$. This is also in accordance with previous observations[22, 24], which is due the fact that the device is still limited by Thouless energy instead of superconducting gap in the long ballistic junction limit as the device length is bigger than the superconducting coherence length $hv_F/\Delta \sim 0.5\mu m$. Note that the measured critical current can also be smaller compared to the intrinsic supercurrent due to unavoidable presence of various form of noises. We further measured the critical current by high frequency differential resistance measurement at 735kHz using the noise measurement setup (Fig. S11), which show larger critical current and better gate voltage independency of the $I_C R_N$ product at higher carrier densities.

Further, contact transparency can be extracted from the normal state resistance $R_N$ in the following way[22, 24]. The Sharvin resistance or the quantum resistance $R_Q$ of the ballistic graphene channel is given by $(h/e^2)4N$, where $N = int(2W/\lambda_F)$ is the number of ballistic modes present, $\lambda_F = 2\pi\sqrt{\pi n}$ is the Fermi wavelength and 'n' is the carrier density [22, 24]. The total resistance of the device is $R_N = R_Q + 2R_C$, where $R_C$ is the contact resistance. Now the contact transmittance is $Tr = R_Q/(R_Q + R_C)$ [22, 24], which gives a mean transmittance of $Tr \sim 0.7$ (Fig. S12) for high electron doping indicating high transparency of the contacts. Note that, in the manuscript we have measured the normal electron transparency or transmittance in QH regime ($T_N$) to be $\sim 0.98$, which is higher compared to the same at zero magnetic field. In a semiclassical picture, in QH regime transport happens via skipping orbits at the QH edges and thus at the QH - Lead interface an incident beam of electron gets transmitted into the Lead via repeated incidences. We believe that due to this repetitive incidence in the QH regime the junction transparency is higher compared to the transparency at zero magnetic field.

**Section S9 | Temperature dependence of Fano factors**

In Fig. S15(a), we plot the shot noise $S_I$ measured at the center of $\nu = 2$ QH plateau at $B = 4T$ as a function of $I_{SD}$ at several temperatures. The corresponding $F_S$ and $F_N$ (from fitting with Eq. (1) of main manuscript) are plotted as a function of temperature in Fig. S15(b). $F_S$ gradually reduces as expected with increasing temperature, and at around $0.5K$ it becomes very close to $F_N$. The enhanced Fano at the lower temperature is definitive proof of AESs, though the deviation from expected Fano of $\sim 0.5$ is due to the normal quasiparticle leakage at higher magnetic field ($\sim 4T$) via the presence of vortices at the QH-superconductor interface. It would have been fantastic if we could have measured the temperature dependence of Fano at lower magnetic field ($\sim 1.5T$), but the smaller QH gap (Fig. S14) restricts us from its study with increasing temperature, which is predominantly dominated by the transport of the normal quasiparticles through the bulk of the sample.

Furthermore, it can be noticed from Fig. S15 that the $F_S$ and $F_N$ become almost equal at $0.5K$ indicating much smaller value of proximity induced superconducting gap than the bulk gap of $MoRe$. In Fig. S16, we present some more data to further support about the reduced value of the proximity induced gap. The Fig. S16 shows shot noise measured near the Dirac point at very small magnetic field $B = 100mT$ (the small magnetic field is applied to kill the tiny supercurrent at the Dirac point), which shows the enhanced noise near zero bias. However, the enhanced noise reduces with increasing temperature and becomes close to high bias Fano at around $\sim 2.9K$ (Fig. S16), indicating only transport via normal quasiparticles. The $2.9K$ is smaller than the $T_C$ ($\sim 8.7K$) of bulk $MoRe$. The measured normal state Fano $\sim 0.4$ (Fig. S16) is consistent with the theoretical predictions[25,26] and experimentally measured Fano of $\sim 1/3$ for graphene at the Dirac point due to pseudodiffusive transport of normal quasiparticles at zero magnetic field[12,27,28]. The slightly higher value than 1/3 in our experiment is possibly due to smaller aspect ratio ($W/L < 4$) as highlighted in Ref. [25]. However, further reduction of proximity induced gap ($0.5K$) in Fig. S15 at $4T$, could be due to further application of magnetic field and need further studies to understand the microscopic details at the interface of QH-superconductor interface.

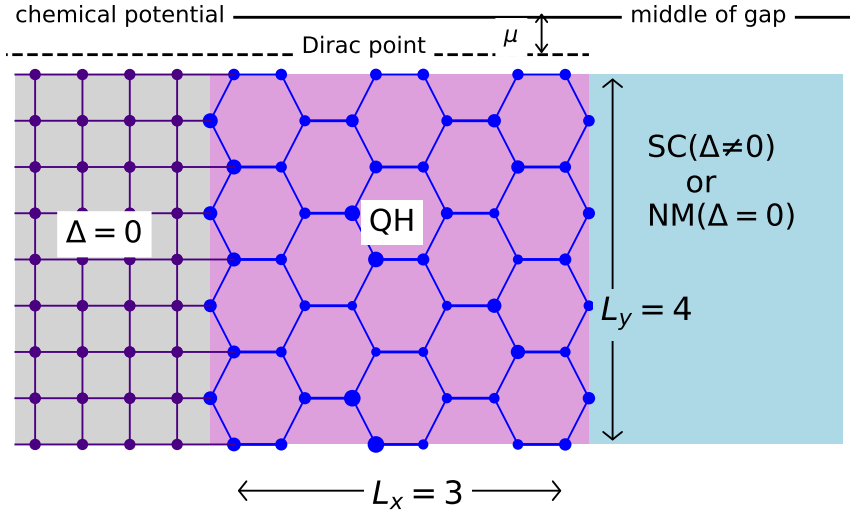

**Figure S1: Schematic of the system in detail.** The left lead is nearest-neighbour tight binding model on a semi-infinite square lattice having width same as the middle region, i.e., of graphene. Graphene is modelled with a nearest-neighbour tight binding hexagonal lattice having magnetic field perpendicular to its plane and onsite disorder. Calculations are done with $L_x = 16$ and $L_y = 60$. The right lead is a tunable region between normal metal (NM) and superconductor (SC) using the gap parameter $\Delta$. The kinetic part in the Hamiltonian of this region has parabolic dispersion. The chemical potential is measured with respect to the Dirac point and always corresponds to the middle of the SC gap.

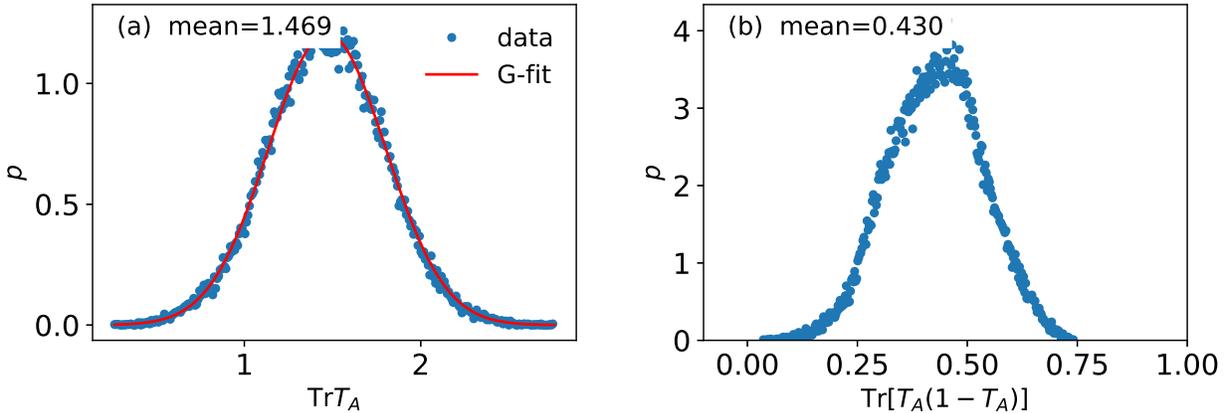

**Figure S2: Disorder distribution of transmission and noise at the $\nu = 6$ plateau.** (a) The distribution $p$ of $\mathrm{Tr}T_A \propto$ conductance ($G$) is shown for the $\nu = 6$ plateau where $\mathrm{Tr}T_A$ has effectively three non-zero eigenvalues coming from three edge channels for each spin. The distribution is close to a Gaussian, with a variance $\sim 0.12$, obtained from a Gaussian (G) fit. The variance is somewhat different from that ($\sim 0.19$) obtained from the distribution of the sum of three uniformly distributed random numbers over $[0, 1]$. (b) Distribution of $\mathrm{Tr}[T_A(1 - T_A)] \propto$ noise ($S$). Number of disorder realization used is 50,000.

9

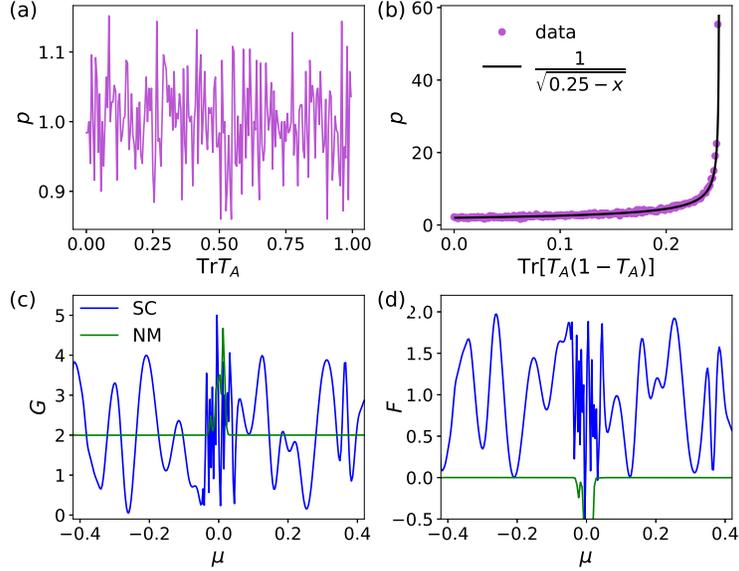

**Figure S3: Disorder distribution of Andreev transmission, and shot noise.** (**a**) Numerically computed probability density $p(\text{Tr}T_A)$ for $\nu = 2$ plateau shows that $\text{Tr}T_A$ is distributed uniformly from 0 to 1 for different disorder realizations in the microscopic model of graphene QH-SC junction. (**b**) Distribution of $x = \text{Tr}[T_A(1 - T_A)]$ that determines shot noise. The uniform distribution of $\text{Tr}(T_A)$ in (**a**) implies a distribution $(0.25 - x)^{-1/2}$, which fits nicely with the distribution computed from the NEGF calculation. This particular distribution gives a mean of 1/6, resulting in the Fano factor of 2/3. The numerically calculated $p(\text{Tr}T_A)$ and $p(\text{Tr}[T_A(1 - T_A)])$ are different from the distributions, $p(x) \sim [x(1 - x)]^{-1/2}$ and $p(x) \sim [x(0.25 - x)]^{-1/2}$, with mean 0.5 and 0.125, respectively, for $P_{AR} \propto \sin^2(\phi/2) \propto \text{Tr}T_A$ and $P_{AR}(1 - P_{AR})$ in the effective toy model. The latter assumes uniform distribution of $\phi$. (**c**) Conductance of the QH-SC junction (blue) as a function chemical potential for a single disorder realization oscillates between 0 and 4 and gives rise to $\nu = 2$ plateau when averaged over many disorder realizations. Green curve is the conductance for the QH-NM case for a single disorder realization. (**d**) The Fano factor for a single realizations (blue) is also oscillatory function of $\mu$. Values of $W$, $\Phi$, and $\Delta$ are chosen to be the same as used in manuscript Fig. 4. The green curve shows the zero Fano factor for the QH-NM case.

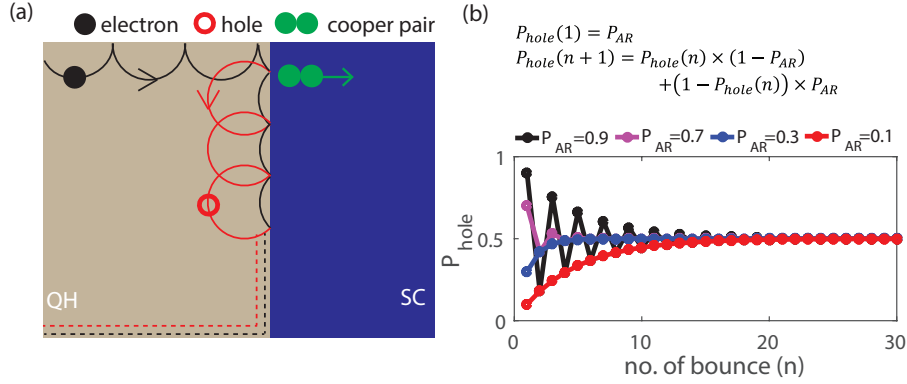

**Figure S4: A semi-classical model.** (**a**) The semi-classical skipping orbit picture of AR at the QH - superconductor junction. (**b**) Top: The recurrence relation relating Probability of getting a hole in the $n$-th bounce to that in the $(n+1)$-th bounce. Bottom: Probability of getting a hole ($P_{hole}$) plotted as a function of number of bounce showing saturation of $P_{hole}$ to half at around 20 bounces. To get a sense of length scale, for graphene 20 bounces at $B=1$ T corresponds to a length of 0.5 $\mu m$

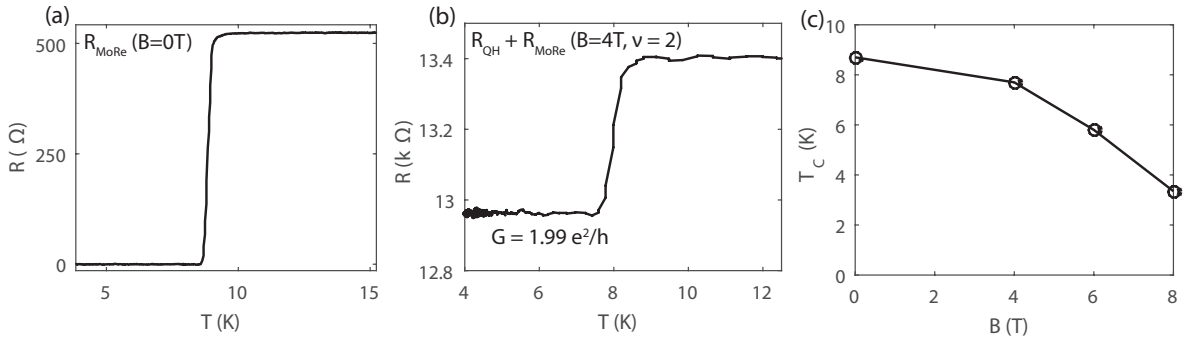

**Figure S5: MoRe characterization.** (**a**) Resistance of the MoRe lead ($R_{MoRe}$)) at $B = 0$ T is plotted as function of temperature showing the superconducting transition around $T_C \sim 8.7$ K. (**b**) Two probe resistance of the device ($R_{QH} + R_{MoRe}$) at $B = 4$ T is plotted as a function of temperature fixing graphene carrier density at the center of $\nu = 2$ plateau. At 4 T superconducting transition is observed around $T_C \sim 7.7$ K. (**c**) $T_C$ is plotted as a function of magnetic field.

11

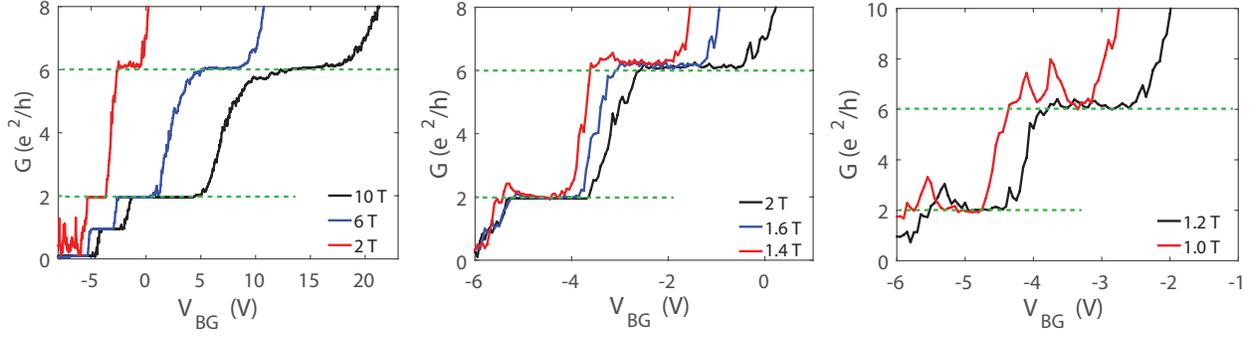

**Figure S6: QH response of the device.** The two probe conductance of the device is plotted as a function of back gate voltage ($V_{BG}$) showing the quantized plateaus at magnetic field as low as 1 T. The conductance values at the QH plateaus of the QH-SC device is same as the value expected with normal contacts.

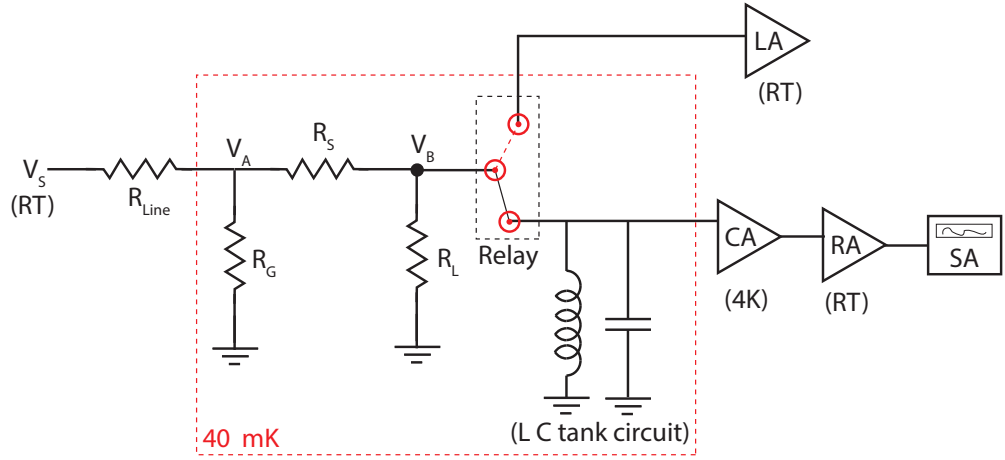

**Figure S7: Measurement setup.** For shot noise measurement we use a LC resonance circuit based high frequency measurement circuit as shown in the schematic. Here line resistance, $R_L = 65\Omega$, resistance to cold ground, $R_G = 5\Omega$, Load resistance, $R_L = 20k\Omega$, $R_S$ (=1/G) is the sample resistance, Inductance , $L \sim 365\mu H$ and capacitance, $C = 128pF$. For low frequency $R_L$ is shunted by the inductor, hence the voltage drop across the sample is $V_{SD} = V_A = V_S \times \frac{R_G}{R_L+R_G}$. Noise is generated due to the applied DC current ($I_{SD} = \frac{V_{SD}}{R_S}$), which is amplified first by a cold amplifier (CA) sitting at the PT2 (4K) plate of the dilution fridge and then by a room temperature (RT) amplifier (RA) and finally measured by a spectrum analyzer (SA). 30 kHz bandwidth (BW) is used for all the high frequency noise measurements. The power spectral density of the output signal ($S_V^{measured}$) is related to the measured voltage ($V_{SA}$) by the relation, $S_V^{measured} = \frac{V_{SA}^2}{BW}$. The required total voltage noise ($S_V$) can be extracted from $S_V^{measured}$ by dividing square of the total gain (g) of the cascade of CA and RA, i.e. $S_V = \frac{S_V^{measured}}{g^2}$. A relay is placed on the Mixing Chamber plate of the dilution fridge to switch the circuit to a lock-in amplifier (LA) for low frequency measurements.

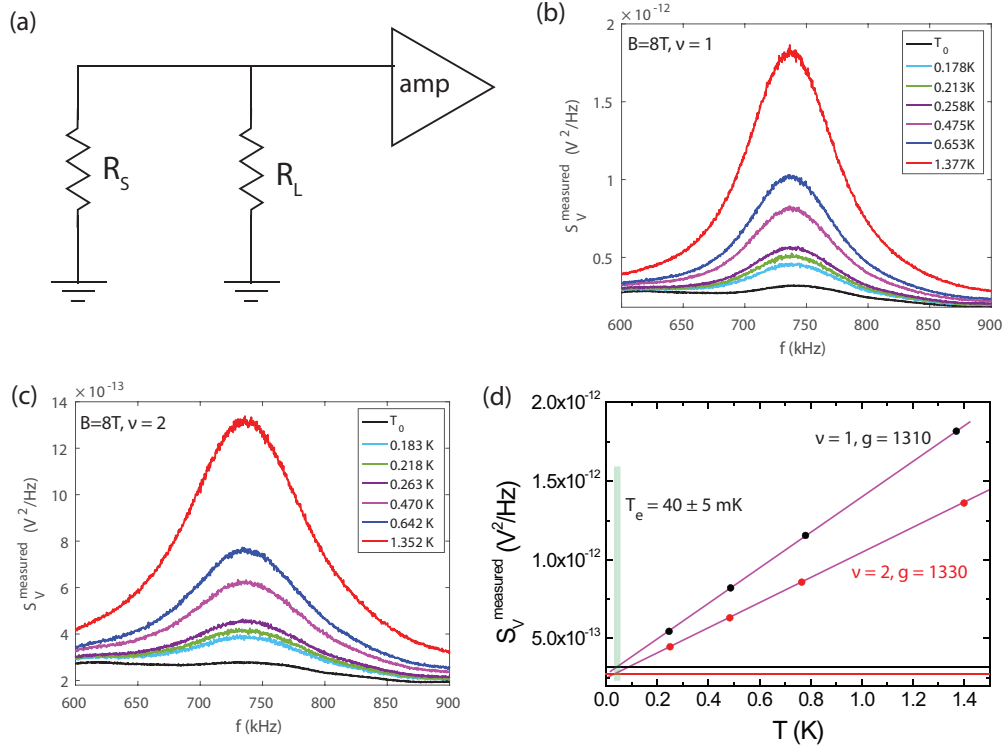

**Figure S8: Gain calibration and electron temperature extraction.** We calculate the gain of cascade of amplifiers (g) by measuring the total noise $S_V^{measured}$ at $I_{SD} = 0$ as a function of temperature. The corresponding effective circuit is shown in **(a)**. In **(b)** and **(c)** we plot $S_V^{measured}$ as a function of frequency for $\nu = 1$ (Black dots) and $\nu = 2$ (red dots) filling factor respectively at B=8T. Here resonance frequency is, $f \sim 735 kHz$, where all the shot noise measurements were performed. The measured signal, $S_V^{measured} = S_V \times g^2 = (S_I R_{eff}^2 + S_i^{amp} R_{eff}^2 + S_v^{amp} + 4k_B T R_{eff}) \times g^2$. Here the last term depends on temperature linearly, and hence $g$ can be extracted from a linear fitting of $S_V^{measured}$ versus T plot as shown in **(d)**. We found the gain to be, $g = 1320 \pm 10$. Further, we can exactly extract electron temperature ($T_e$) from the fitting, by looking at the temperature at which the fitted line gives the value of the measured total noise (at $I_{SD} = 0$) at base temperature of the fridge. Black and red horizontal lines, respectively, show the base temperature noise signal $S_V^{measured}$ (at $I_{SD} = 0$) for $\nu = 1$ and $\nu = 2$ filling factor. We found $T_e$ to be $40 mK \pm 5 mK$ (shaded vertical green region marks the approximate error).

13

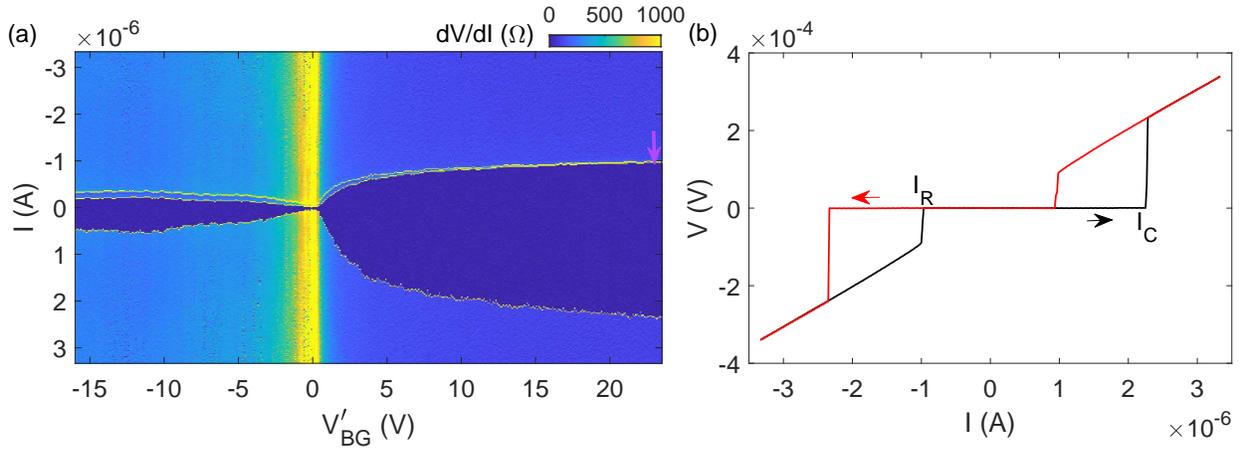

**Figure S9: Supercurrent at zero magnetic field.** (a) 2D colormap of the differential resistance $dV/dI$ plotted as a function of gate voltage relative to Dirac point ($V'_{BG} = V_{BG} - V_{DP}$) and applie DC current (I) showing supercurrent at all carrier densities. (b) Supercurrent at $V'_{BG} = 23.5V$ (magenta arrow in (a)) showing the critical current $I_C$ and the re-trapping current $I_R$ for the two different sweep directions.

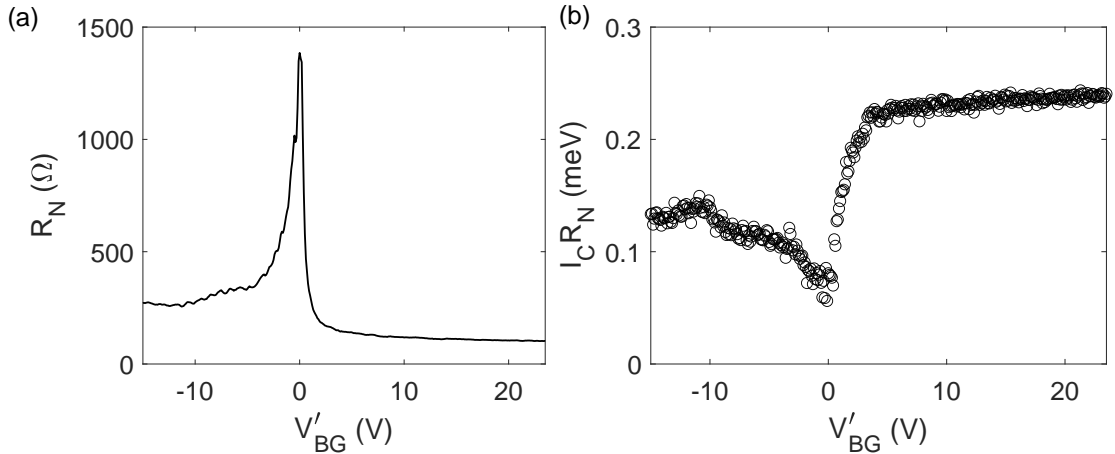

**Figure S10: The $I_C R_N$ product.** (a) Normal state resistance ($R_N$) plotted as a function of $V'_{BG}$. (b) $I_C R_N$ plotted as a function of $V'_{BG}$.

14

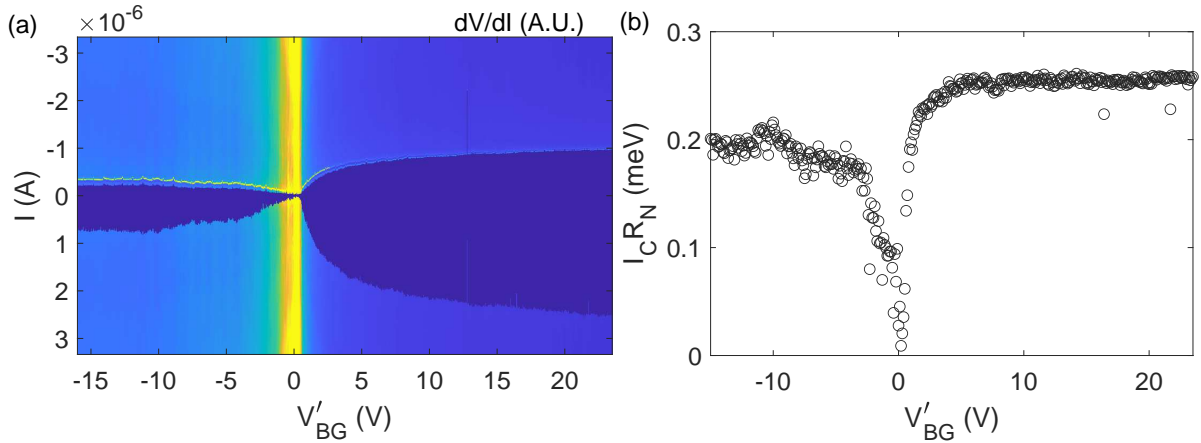

**Figure S11: Critical current measured at 735 kHz.** (**a**) 2D colormap of the differential resistance measured at 735kHz using the high frequency noise measurement setup plotted as a function of $V'_{BG}$ and applied DC current showing larger supercurrent compared to DC measurement (shown in Fig. S15). (**b**) The corresponding $I_C R_N$ product plotted as a function of showing its larger magnitude. Here it is quite apparent that the $I_C R_N$ product does not depend much on the $V'_{BG}$ at large carrier densities.

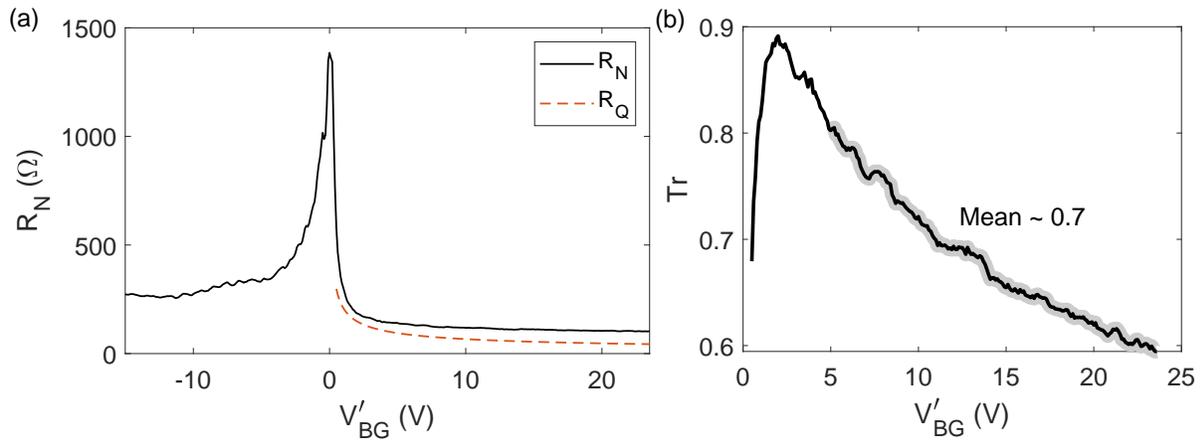

**Figure S12: Extraction of average normal electron transmittance of Graphene-MoRe junction.** (**a**) Sharvin resistance or quantum resistance ($R_Q$) is plotted together with RN as a function of $V'_{BG}$. (**b**) The extracted transmittance ($Tr$) as a function of $V'_{BG}$. Mean transmittance $Tr \sim 0.7$ is extracted from the range where $I_C R_N$ remain independent of $V'_{BG}$.

15

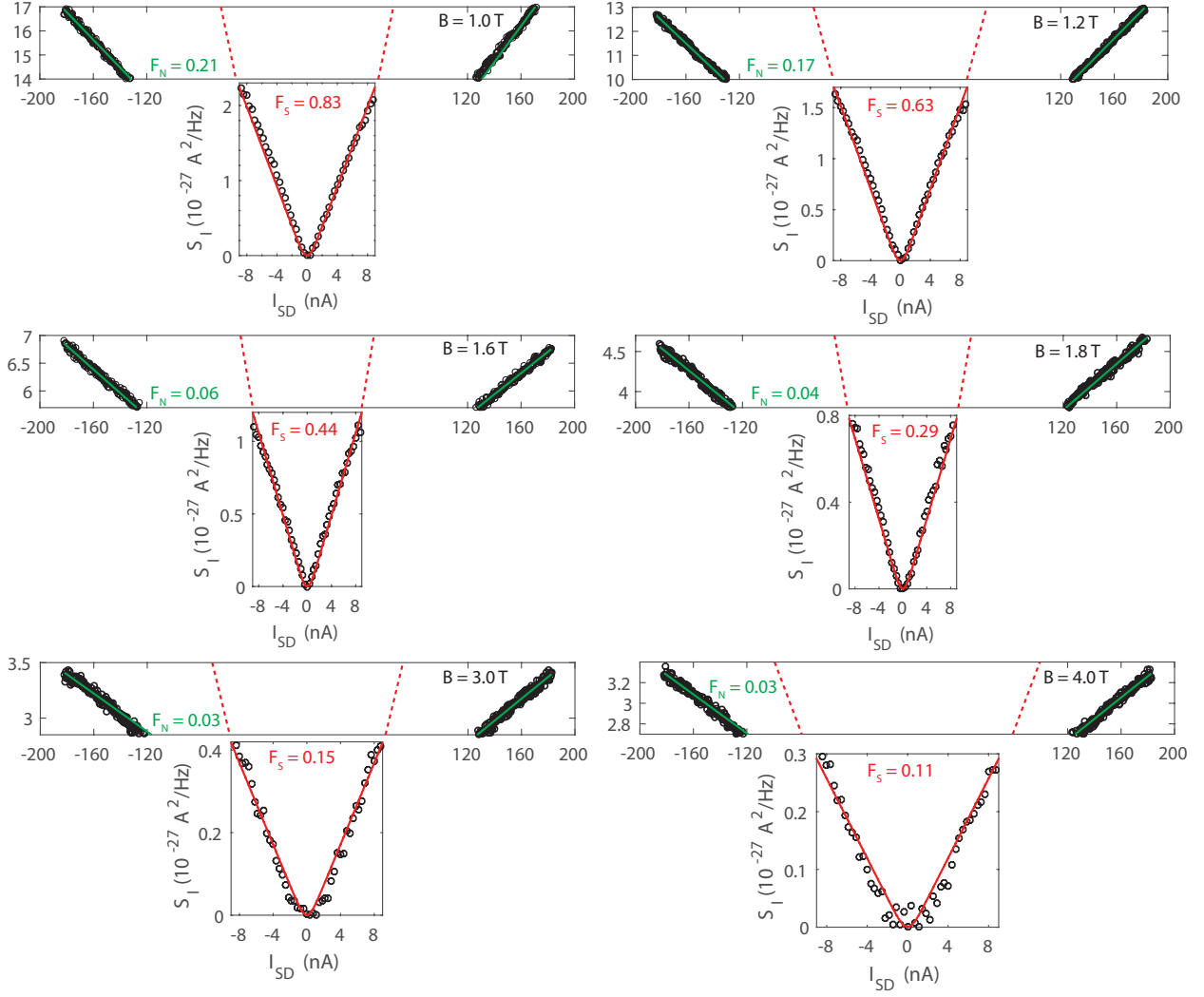

**Figure S13: Additional shot noise data at $\nu = 2$ filling factor at different B.** Measured shot noise ($S_I$) is plotted as a function of applied bias current ($I_{SD}$) at several magnetic fields. The low bias ($eV_{SD} < 2\Delta$) and high bias ($eV_{SD} > 2\Delta$) shot noise are fitted using Eq. (1) of main manuscript to extract the corresponding Fano factors $F_N$ and $F_S$.

16

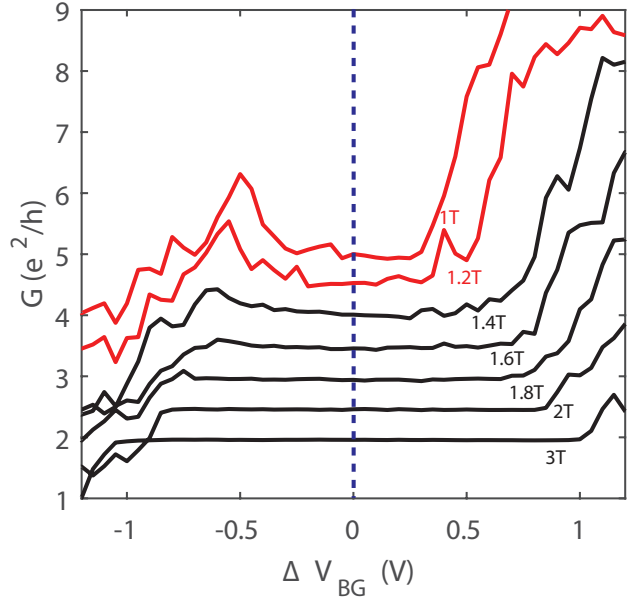

**Figure S14: Quality of conductance plateau at $\nu = 2$ filling factor.** Conductance at the $\nu = 2$ plateau is plotted as a function of $\Delta V_{BG}$ (shifted such that the center of $\nu = 2$ plateau appears at $\Delta V_{BG} = 0$), showing the degrading plateau quality above 1.4T in support of our claim of increasing bulk contribution to shot noise in this regime. Below 3T the conductance is gradually shifted by $0.5e^2/h$ for each data.

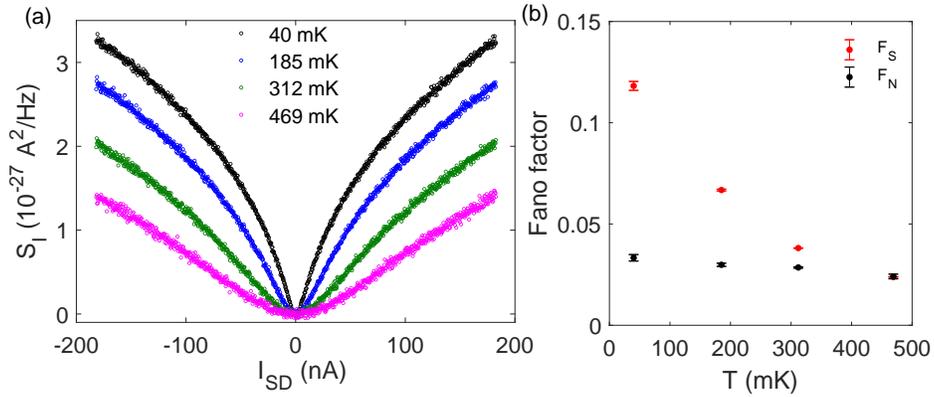

**Figure S15: Temperature dependence of Fano factors.** (**a**) $S_I$ plotted as a function of applied DC curent ($I_{SD}$) at the center of $\nu$=2 QH plateau at B=4T. (**b**) $F_N$ and $F_S$ plotted as a function of temperature.

17

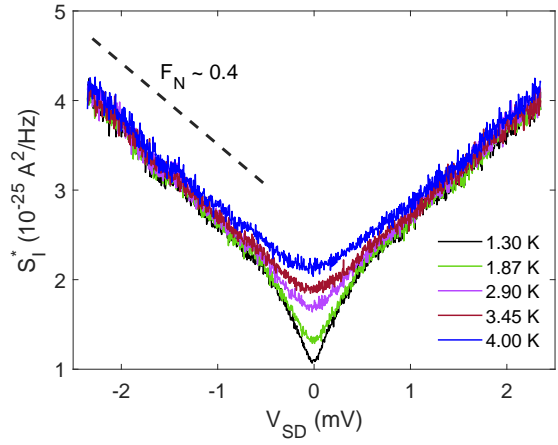

**Figure S16: Enhanced noise at B=100mT.** The total current noise $S_I^*$ ($=S_V/R_{eff}^2 = S_I + 4k_BT/R_{eff} + S_i^{amp} + S_v^{amp}/R_{eff}^2$) plotted as a function of applied DC voltage ($V_{SD} = I_{SD}/G$) near the Dirac point at B=100mT. Enhancement in shot noise is observed near zero bias which goes away around $\sim 2.9K$. We plot $S_I^*$, which is simply $S_I$ combined with current independent thermal noise and amplifier noises, to better visualize $F_S$ approaching $F_N$ at around $2.9K$.



\end{document}